\documentstyle[pre,twocolumn,aps,psfig]{revtex}
\parskip=\medskipamount  %
\begin{document}
\preprint{}            %
\input{epsf}           %

\twocolumn[\hsize\textwidth\columnwidth\hsize\csname @twocolumnfalse\endcsname  %
 
\title{\large \bf The Use of Generalized Information Dimension
in Measuring Fractal Dimension of Time Series}
\author{Y. Ashkenazy}
\address{Department of Physics, Bar-Ilan University, Ramat-Gan 52900, Israel}

\date{\today}
\maketitle

\begin{abstract}
{An algorithm for calculating generalized fractal dimension of a
time series using the
general information function is presented. The algorithm is based on a strings
sort technique and requires $O(N \log_2 N)$ computations. A rough estimate
for the number of points needed for the fractal dimension calculation is given.
The algorithm was tested on analytic example as well as well-known examples,
such as, the Lorenz attractor, the Rossler attractor, the van der Pol 
oscillator, and the
Mackey-Glass equation, and compared, successfully, with previous results 
published in the literature. The computation time for the algorithm 
suggested in this paper is much less then the computation time according to 
other methods.}
\end{abstract}

\vskip2pc]  %
\narrowtext

\vspace*{-2.0truecm}   %

\section{Background} \label{s1}

In the recent decades the study of chaos theory has gathered momentum.
The complexity that can be found in many physical and biological systems
has been analyzed by the tools of chaos theory. 
Characteristic properties, such as the Lyapunov
exponent, Kolmogorov entropy and the fractal dimension (FD), have been
measured in experimental systems. It is fairly easy to calculate signs of
chaos if the system can be represented by a set of non-linear ordinary 
differential equations. In many cases it is very difficult to built a
mathematical model that can represent sharply the experimental system. 
It is essential, for this purpose, to reconstruct a new phase space based on 
the information that one can produce from the system. A global value that is 
relatively simple to compute is the FD. The FD can give
an indication of the dimensionality and complexity of the system.
 Since actual living biological systems 
are not stable and the system complexity varies with time,
one can distinguish between different states of the system by the FD. 
The FD can also determine whether a particular system is more complex than
other systems. However, since biological systems are very complex, it is better
to use all the information and details the system can provide. In this paper 
we will present
an algorithm for calculating FD based on the geometrical structure
of the system. The method can provide important information, in addition, on 
the geometrical of the system (as reconstructed from a time series).

The most common way of calculating FD is through the correlation function,
$C_q(r)$ (eq. \ref{e7}). There is also other method of FD calculation based 
on Lyapunov exponents and Kaplan-Yorke conjecture \cite{Kaplan79} 
(eq. \ref{e24}). However, the computation of the Lyapunov exponent spectrum 
from a time series is very difficult and people usually try to avoid this 
method\footnote{The basic difficulty is that for high FD there are some 
exponents which are very close to zero; one can easily add an extra 
unnecessary exponent
that can increase the dimensionality by one; this difficulty is most
dominant in Lyapunov exponents which have been calculated from a time series.}.
The algorithm which is presented in this paper is important, since it gives a 
comparative method for calculating FD according to the correlation function. 
The need for a additional method of FD calculation is critical in some types 
of time series, 
such as EEG series (which are produced by brain activity), since several 
different
estimations for FD were published in the literature 
\cite{{Babloyantz85},{Basar90},{MayerKress88},{Saermark89},{Saermark97}}. 
A comparative algorithm
can help to reach final conclusions about FD estimate for the signal.

A very simple way to reconstruct a phase space from single time series
was suggested by Takens \cite{Takens81}. Giving a time series ${x_i}, 
i=1\ldots N_p$, we build a new $n$ dimensional phase space in the 
following way :
\begin{equation} \label{e1}
\left.
\begin{array}{l}
\vec y_0 = \{x(t_0),x(t_0+\tau ),x(t_0+2\tau ),\ldots
,x(t_0+(n-1)\tau )\} \\
\vec y_1 = \{x(t_1),x(t_1+\tau ),x(t_1+2\tau ),\ldots
,x(t_1+(n-1)\tau )\} \\
\vdots\\
\vec y_{N} = \{x(t_{N}),x(t_{N}+\tau ),x(t_{N}+2\tau
),\ldots ,x(t_{N}+(n-1)\tau )\} \\
\\
t_i=t_0+i\Delta t \qquad \tau =m\Delta t \\ 
N=N_p-(n-1)m \qquad m=integer,
\end{array}
\right.
\end{equation}
where $\Delta t$ is the sampling rate, $\tau$ corresponds to the interval  
on the time series
that creates the reconstructed phase space (it is usually chosen to be the 
first zero of the autocorrelation function \cite{Sch89}, or the first minimum
of mutual information \cite{Abarbanel93};
 in this work we will use $m$ instead of $\tau$ as an index), and $N$ is 
number of reconstructed vectors.
For ideal systems (an infinite number of points without external noise) 
any $\tau$ can
be chosen. Takens proved that such systems converge to the real dimension
if $n \geq 2D+1$, where $D$ is the FD of the real system. According
to Ding {\it et al} \cite{Ding93} if $n \geq D$ then the FD of the 
reconstructed phase space is equal to the actual FD.

In this paper we will first (sec. \ref{s2}) present regular analytic 
methods for calculating the FD of a system (generalized correlation method and 
generalized information method). An efficient
algorithm for calculating the FD from a time series based on string sorting
will be described in the next section (sec. \ref{s3}). The following
 step is to check and compare the general correlation
methods with the general information method (see eq. (\ref{e2}))
using known examples (sec. \ref{s4}). 
Finally, we summarize in sec. \ref{s6}.

\section {Generalized information dimension and
generalized correlation dimension} \label{s2}

\subsection {Generalized information dimension}
The basic way to calculate an FD is with the Shannon entropy, 
$I_1(\varepsilon)$\footnote{Generally, one has to add a superscript, $n$, 
the embedding 
dimension in which the general information is calculated. At this stage we
assume that the embedding dimension is known.} \cite{Shannon49}, of 
the system. The entropy is just a particular case of the general information 
which is defined in the following way \cite{Bala76} :
\begin{equation}
I_q(\varepsilon)=\frac{1}{1-q}\ln(\sum_{i=1}^{M(\varepsilon)} p_i^q), 
\label{e2}
\end{equation}
where we divide the phase space to $M(\varepsilon)$ hypercubes of edge 
$\varepsilon$. The probability to find a point in the $i^{th}$
hyper cube is denoted by $p_i$. The generalization, $q$, can be any 
real number; we usually
use just an integer $q$. When we increase $q$, we give more 
weight to more populated
boxes, and when we decrease $q$ the less occupied  boxes become dominant.
For $q=1$, by applying the l'Hospital rule, eq. (\ref{e2}) becomes :
\begin{equation}
I_1(\varepsilon) = -\sum_{i=1}^{M(\varepsilon)} p_i \ln p_i,  \label{e3}
\end{equation}
where $I_1(\varepsilon)$ is referred to also as the Shannon entropy 
\cite{Shannon49} of the system.
The definition of the general information dimension (GID) is :
\begin{equation}
D_q = -\lim_{\varepsilon \to 0} \frac {I_q(\varepsilon)}{\ln \varepsilon}, 
\label{e4}
\end{equation}
for $N \to \infty$ ($N$ is the number of points). 
However, in practice, the requirement ${\varepsilon \to 0}$ is not 
achieved, and an average over a range of $\varepsilon$ is required. 
\begin{figure}[thb]
\psfig{figure=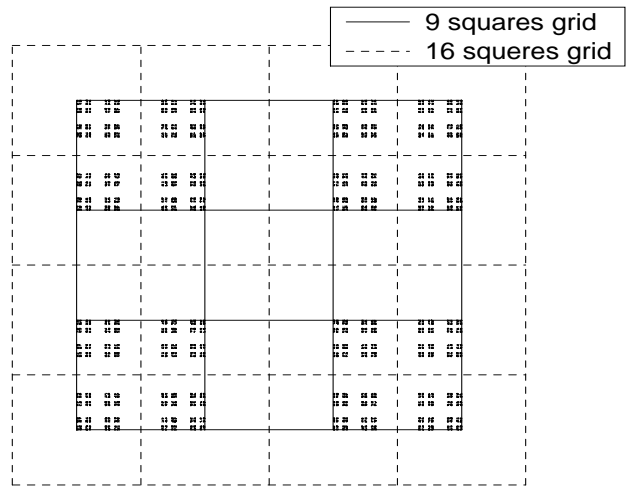,height=8cm,width=10cm,angle=-90}
\caption[]{\label{fig1}$2D$ Cantor set. Two different positions of the 
two dimensional grid, give completely different FD. }
\end{figure}

In some cases this average is not sufficient
 because several values of $I_q(\varepsilon)$
can be computed for the same $\varepsilon$. To illustrate this 
we use a $2D$ Cantor
set, presented in Fig. \ref{fig1}. We start from a square. From 9 sub-squares 
we erase the 5 internal squares. We continue with this evolution for each
remainder square. This procedure is continued, in principle,
 to infinity. The FD, $D_0$, of the 
$2D$ Cantor set is $\ln 4 / \ln 3$ (when $q=0$, $I_0$ is just the number of
nonempty squares and $D_0$ is the logarithmic ratio between nonempty
squares and $\varepsilon$, where the square edge is normalized to one) as shown
in Fig \ref{fig1}. However, it is possible to locate the $2D$ grid in such 
a way that there are 16 nonempty squares, giving rise to 
$D_0 = 2 \ln 4 / \ln 3$, twice the time of real FD of the system. This 
illustration shows that when $\varepsilon$ is not small, different 
positions of the grid can lead to different FD's. In this case it is clear
that we have to locate the grid in such a way 
that minimum squares will be nonempty
(it is easy to show that this claim for the minimum is true for every $q$).
In general, we can say that one must locate the hyper-grid so that 
the general information is minimum :
\begin{equation}
\tilde{I_q}(\varepsilon) \equiv \min 
\frac{1}{1-q}\ln(\sum_{i=1}^{M(\varepsilon)} p_i^q), \label{e5}
\end{equation}
This proper location of the hyper-grid reduces the influence of surface 
boxes that are partly contained in the attractor. The requirement 
of a minimum
gives a good estimate for the GID when $\varepsilon$ is not small.

\subsection {Generalized correlation dimension}
In 1983 Grassberger and Procaccia presented a new method for 
calculating $D_2$ \cite{Gras83}. According to this method, it is possible to 
calculate dimension just from the correlation between different 
points, without direct connection to the phase space, and therefore
easy to use. Some years later, a more general correlation function 
 was suggested by Pawelzik and Schuster \cite{Pawe87} : 
\begin{equation}
C_q(r) = \Biggl[ \frac{1}{N} \sum_{i=1}^{N} \biggl[ \frac{1}{N}
\sum_{j=1}^{N} \Theta \bigl( r - \left| \vec x_i - \vec x_j \right| 
\bigr) \biggr]^{q-1} \Biggr]^{1 \over q-1},  \label{e7}
\end{equation}
where $N$ is number of points, $\vec x_i$ is a point of the system
and $q$ is the generalization. $\Theta (x)$ is the Heaviside step function
\begin{equation}
\Theta (x) = \left\{
  \begin{array}{ll}
     0   & {\rm when~} x \leq 0\\
     1   & {\rm when~} x > 0. 
  \end{array} \right. \label{e8}
\end{equation}
According to this method we have to calculate the generalized 
probability to find any two points within a distance $r$. This method is
some kind of integration on eq. (\ref {e2}). It is not necessary to compute 
the real distance (e.g. $\left| \left| \vec x \right| \right| 
= \sqrt{x_1^2 + \cdots +x_n^2}$ 
where n is the phase space dimension); it is equivalent to
calculating the probability to find any two points in a hyper-box where one
of them is located in the middle (e.g. $\left| \left| \vec x \right|
\right| = \max_{1 \le i \le n} \left| x_i \right| $ \cite{Gras90}). 
It is easier to compute the last possibility. For the special case
of $q=1$, eq. (\ref{e7}) can be written (by applying the l'Hospital's rule
\cite{Abarbanel93}) as :
\begin{equation}
\ln C_1(r) = \frac{1}{N} \sum_{i=1}^{N} \ln \biggl[ \frac{1}{N} 
\sum_{j=1}^{N} \Theta \bigl( r - \left| \vec x_i - \vec x_j \right| 
\bigr) \biggr].  \label{e8a}
\end{equation}

The generalized correlation dimension (GCD) has a similar definition to
the GID (eq. (\ref {e4})) :
\begin{equation}
D_q = \lim_{N \to \infty ; r \to 0} \frac{\ln C_q(r)}{\ln r}. \label{e9}
\end{equation}
Both GID and GCD methods should give identical results.
The GCD method is easy use and gives smooth curves. On the other hand,
the method requires $O(N^2)$ computations\footnote{If one looks just at small 
$r$ values, the method requires just $O(N)$ computations \cite{Gras90}.}. 
Also, the smooth curves due to
averaging over all distances are associated with a loss in information based 
on the attractor structure.

As we pointed out earlier, we usually have a limited number of points,
forcing us to calculate dimension at large $r$. Thus an error enters the
calculation of FD. The minimum number of points needed for the FD 
calculation has been discussed in several papers and different constraints 
suggested (for example \cite{Smit88} and \cite{Eckm92}). In this paper
we will use the $N_{min}$ of the Eckmann and Ruelle \cite{Eckm92} constraint :
\begin{equation}
D_2 < \frac{2 \log_{10} N}{\log_{10} \Bigl( \frac{1}{\rho} \Bigr) },
 \label{e10}
\end{equation}
under the following conditions :
\begin{equation}
\rho = \frac{r}{r_0} \ll 1 \qquad \frac{1}{2} N^2 \rho^D \gg 1,  \label{e11}
\end{equation}
with reference to Grassberger and Procaccia method. Here, $r_0$ is the 
attractor diameter, and
$r$ is the maximum distance that gives reliable results in eq. (\ref{e7}) when
$q=2$. The normalized distance, $\rho$, must be small because 
of the misleading influence of the hyper-surface. 
A $\rho$ too large (close to 1) can cause incorrect results since we
take into account also hyper-boxes that are not
well occupied because the major volume is outside the attractor.

However, if one has a long time series, then it is not necessary to compute
all $N^2$ relations in eq. (\ref{e7}). One can compute eq. (\ref{e7}) for 
certain reference points such that the conditions in eq. (\ref{e11}) will 
still hold. Then, eq. (\ref{e7}) can be written as follows :
\begin{eqnarray}
\label{e12}
C_q(r) &=& \Biggl[ \frac{1}{N_{ref}} \sum_{i=1}^{N_{ref}} \biggl[ 
\frac{1}{N_{data} - 2w -1} \times \nonumber \\
&& \sum_{j=1}^{N_{data}} \Theta 
\bigl( r - \left| \vec x_{i \cdot s} - \vec x_j \right| 
\bigr) \biggr]^{q-1} \Biggr]^{1 \over q-1},
\end{eqnarray}
where
$$
\left| i \cdot s - j \right| > w \qquad s = {\lfloor {\frac{N_{data}}{N_{ref}}
}\rfloor}.
$$
For each reference point we calculate the correlation function over all data
points. The step for the time series is $s$. To neglect short 
time correlation one must also introduce a cutoff, $w$. Usually
$w \approx m$ ($m$ corresponds to the $\tau$ from (\ref{e1}))
 \cite{Thei86}. The number of distances $\left| \vec x_i - \vec x_j 
\right|$ will be $P = N_{ref}(N_{data} -2w -1)$ instead of $N^2$. The new form
 of eq. (\ref{e10}) is:
\begin{equation}
D < \frac{\log_{10} P}{\log_{10} \Bigl( \frac{1}{\rho} \Bigr) }, \label{e13}
\end{equation}
and the conditions (\ref{e11}) become
\begin{equation}
\rho = \frac{r}{r_0} \ll 1 \qquad \frac{1}{2} P \rho^D \gg 1.  \label{e14}
\end{equation}
Although we discussed here a minimum number 
of data points needed for the dimension
calculation, one must choose $\rho$ such that the influence from
the surface is negligible (the growth of the surface is exponential
to the embedding dimension). That gives us an upper limit for $\rho$.
On the other hand, in order to find the FD we must average over a range of
$r$, giving us a lower limit for $\rho$; therefore $N_{min}$
is determined according to the lower limit, giving rise to larger amount of
points.

\section {Algorithm} \label{s3}
\subsection {Background}
In this section we describe an algorithm for GID method\footnote{Computer 
programs are available at 
\it{http://faculty.biu.ac.il/}$\sim$\it{ashkenaz/FDprog/}}. 
The algorithm is based on a string sort and can be useful
both for GID and GCD methods.

In previous works there have been several suggestions to compute DIG 
\cite{{Casw86},{Molteno93},{Hou90},{Block90},{Lieb89},{Kruger96}}. The key
idea of some of those methods \cite{{Hou90},{Block90},{Lieb89}} is to rescale
the coordinates of each point and to express them in binary form. Then, the
edge-box size can be initialized by the lowest value possible and then 
double it each time. Those methods require $O(N \log N)$ operations. Another 
method \cite{Molteno93} uses a recursive algorithm to find the FD. The 
algorithm
starts from a $d$ dimensional box which contains all data points. Then, this
box is divided into $2^d$ sub boxes, and so on. The empty boxes are not 
considered, and those which contain only one point are marked. This procedure
requires only $O(N)$ operations (for a reasonable series length this fact does 
not make a significant difference \cite{Molteno93}).

As pointed out in ref. \cite{Molteno93}, in spite of the efficiency of the 
above algorithm (speed and resolution), it is quite difficult to converge to 
the FD of a high dimensional system. We were aware to this difficulty, and 
suggested a {\it smoothing} term to solve it (as will be explained in this 
section). Basically, we allow any choice of edge length (in contrast to power
of 2 edge size of the above methods) and optimal location of the grid is 
searched. This procedure leads to convergence to the FD.

One of the works that calculates GID was done by Caswell and Yorke
\cite{Casw86}. They calculate FD of $2D$ maps.
They have proved that it is very efficient to 
divide the phase space into circles instead of squares in spite of the 
neglected area between the circles. However, this 
approach is not suitable for higher phase space dimensions. The reason might
be that 
the volume of the hyper-balls compared to the entire volume of the attractor 
decreases according to the attractor dimension,
and in this way we lose most of the information that is included in the
attractor, since the space between the hyper-balls is not taken into account. 
It is easy to show 
that the ratio between the volume of the hyper-sphere with radius 
$R$, $V_S$, and the volume of the hyper-box with edge size of 
$2R$, $V_B$ (the sphere is in the 
box), is
\begin{equation} \label{e14a}
\frac{V_S(2n)}{V_B(2n)} = \frac{1}{2\cdot4\cdot6\cdots2n} \Bigl(
\frac{\pi}{2}\Bigr)^n,
\end{equation}
for even phase space dimension, and,
\begin{equation} \label{e14b}
\frac{V_S(2n-1)}{V_B(2n-1)} = \frac{1}{1\cdot3\cdot5\cdots(2n-1)} 
\Bigl(\frac{\pi}{2}\Bigr)^{n-1},
\end{equation}
for odd phase space dimension. It is clear that the ratios (\ref{e14a}) and 
(\ref{e14b}) tend to zero for large $n$.

\subsection {Algorithm}
Let us call the time series $x(i)$. The form of
the reconstructed vector, (\ref{e1}),
depends on the jumping, $m$, that creates the phase 
space. We order the time series in the following way :
\begin{equation} \label{e15}
\left.
\begin{array}{l}
x(0),x(m),x(2m),\ldots,x((l-1)m), \\
x(1),x(1+m),x(1+2m),\ldots,x(1+(l-1)m),\\
\vdots\\
x(m-1),x(m-1+m),\ldots,x(m-1+(l-1)m),\\
\end{array}
\right.
\end{equation}
where $l = \lfloor \frac{N_p}{m} \rfloor$. Let us denote the new series 
as $\tilde x_i$
($i = 0 \ldots (lm-1)$; we lost $N_p-lm$ data points). If we take 
$n$ consecutive numbers in each row, we create a reconstructed vector 
in the $n$ dimensional embedding dimension.

At this stage, we fit a string (for each $\varepsilon$) to the 
number series (\ref{e15}). One character is actually a number between 
$0 \ldots 255$, 
and thus, it is possible to divide one edge of the hyper-box to 255 parts 
(we need one of the 256 characters for sign). The correspondence is
applied in the following way. We search for the minimum value in $\tilde x_i$
series, and denote it as $\tilde x_{min}$ (similarly, we denote the maximum 
values as $\tilde x_{max}$).
The corresponding character to $\tilde x_i$ is :
\begin{equation} \label{e16}
y_i = \lfloor \frac{\tilde x_i - \tilde x_{min}}{\varepsilon} \rfloor.
\end{equation}
The value of $\varepsilon$ is in the range
\begin{equation} \label{e16a}
\frac{(\tilde x_{max} - \tilde x_{min})}{255} \le \varepsilon
\le (\tilde x_{max} - \tilde x_{min}).
\end{equation}
If we take now $n$ consecutive characters (string), we have the ``address''
of the hypercube in which the corresponding vector is found.

Let us represent the string as follows :
\begin{equation} \label{e17}
\left.
\begin{array}{l}
s_0 = y_0 y_1 \ldots y_{n-1}, s_1 = y_1 \ldots y_{n},\ldots,
s_{l-n} = y_{l-n} \ldots y_{l-1}, \\
s_{l-n+1} = y_l \ldots y_{l+n-1},\ldots,
s_{2(l-n)+1} = y_{2l-n} \ldots y_{2l-1}, \\
\vdots\\
s_{(m-1)(l-n+1)} = y_{(m-1)l} \ldots y_{(m-1)l+n-1}, \ldots, \\
\qquad \qquad \qquad \qquad \qquad s_{m(l-n+1)-1} = y_{ml-n} \ldots y_{ml-1}.\\
\end{array}
\right.
\end{equation}
Obviously, we do not have to keep each string $s_i$ in the memory; we can keep 
just the pointers to the beginning of the strings in the characters series,
$y_i$. The number of vectors / strings is $m(l-n+1)$.

As mentioned, each string is actually the address of a hyper-box (which the 
vector is in it) in a hyper-grid that cover the attractor. The first 
character is the position of the first coordinate of the hyper-box on the
first edge of the hyper-grid. The second character 
denotes the location on the second edge, and so on. We actually grid the 
attractor with a hyper-grid so that any edge in this grid can be divided into
255 parts. Thus the maximal number of boxes is $255^n$, where $n$ is the 
embedding dimension (there is no limit for $n$). Most of those boxes are 
empty, and we keep just the occupied boxes in the hyper-grid.
\begin{figure}[thb]
\psfig{figure=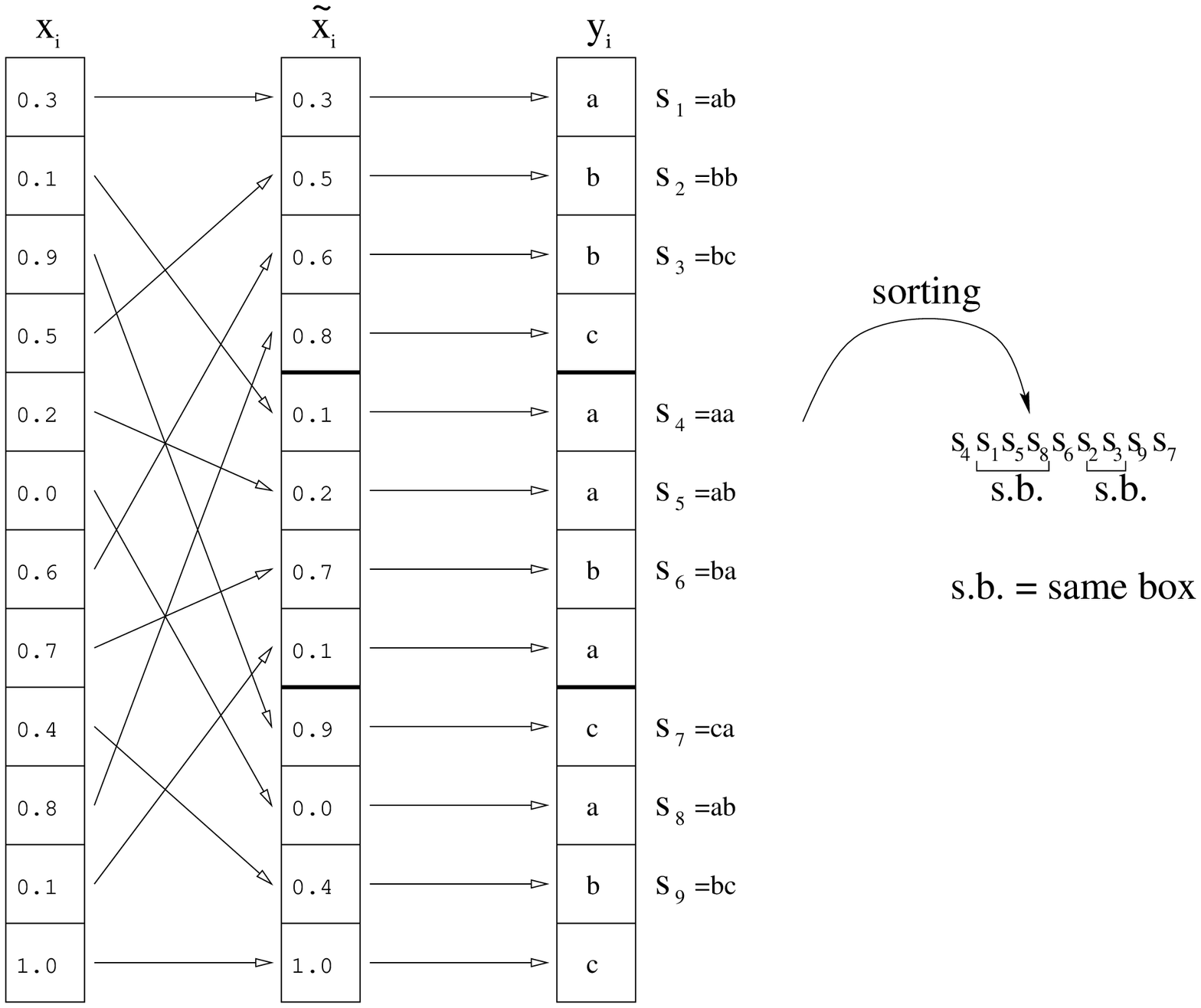,height=8cm,width=8.5cm,angle=-90}
\caption[]{\label{figN2}
Illustration of the GID algorithm. We take series containing $12$ data points
between zero and one. The parameter values are: $m=3$, $\varepsilon=0.4$ and
$n=2$. For simplicity we assume that the lowest character is 'a'.}
\end{figure}

The next step is to check how many vectors fall in the same box, or, in other
words, how many identical ``addresses'' there are. For this we have to sort
the vector that points to strings, $s_i$, in increasing order (one string is
less then other when the first character that is not equal is less then the 
parallel character; e.g., 'abcce'$<$'abcda' since the fourth character in the 
first string, c, is less then the fourth character in the second string, d).
The above process is illustrated in Fig. \ref{figN2}.

The most efficient way to sort $N$ elements is the ``quick sort'' 
\cite{{nr95},{Sort}}. It requires just $O ( N \log_2 N )$ computations.
However, this sort algorithm is not suitable for our propose, since after the
vector is sorted, and we slightly increase the size of the edge of the 
hyper-box, $\varepsilon$, 
there are just a few changes in the vector, and most of it remains
sorted. Thus, one has to use another method which requires less computations
for this kind of situation, because the quick sort requires 
$O ( N \log_2 N )$ computations independently of the initial state. The sort 
that we used was a ``shell sort'' \cite{nr95}, which requires $O ( N^{3/2} )$ 
computations in worst case, around $O ( N^{5/4} )$ computations for random 
series, and $O(N)$ operations for an almost sorted vector. 
Thus, if we compute the information function for $m$ different
$\varepsilon$ values $O( N \log_2 N + mN )$ operations will be needed.

After sorting we count how many identical vectors there are in each box.
Suppose that we want to calculate GID for embedding dimensions
$n_{min} \ldots n_{max}$, and generalizations $q_{min} \ldots q_{max}$. 
We count in
the sorted pointers vector identical strings containing $n_{max}$ characters.
Once we detect a difference we know that the coordinate of the box 
has changed. We detect the first different location. First 
difference in the last character of the string means that it is possible
to observe the difference just in the $n_{max}^{th}$ embedding dimension, and 
in lower
embedding dimension it is impossible to observe the difference. If the
first mismatch is in the $n^{th}$ character in the string, then the box 
changes for embedding 
dimensions higher than or equal to $n$. We hold a counter vector for the 
different embedding dimensions, and we will assign zero for those embedding
dimensions in which we observed a change.
In this stage, we have to add to the results matrix the probability to fall in
a box according to eqs. (\ref{e2}), (\ref{e3}). We continue with this method
to the end of the pointer vector and then print the results matrix for 
the current $\varepsilon$ and continue to the next $\varepsilon$.

It is easy to show that one does not have to worry about the range of the 
generalization, $q$, because for large time series with $10^6$ data
points the range of computational $q$'s is -100..100.

The method of string sorting can be used also for calculating GCD. Our 
algorithm is actually a generalization of the method that was suggested by
Grassberger \cite{Gras90}. The algorithm is specially efficient for small
$r$ ($r< \frac{1}{10}D$ where $D$ is the attractor diameter) and it requires
$O(N\log_2 N)$ computations instead of $O(N^2)$ computations (according to the
Grassbeger algorithm it requires $O(N)$ computations instead $O(N^2)$
computations).

The basic idea of Grassberger was that for small $r_{max}$ values, where
$r_{max}$ is the maximum distance for which the correlation function is 
computed, one does not have 
to consider distances larger then $r_{max}$, distances which require most of 
the computation time. Thus, it is enough to consider just the neighboring 
boxes (we use the definition distance $\left| \left| \vec x \right|
\right| = \max_{1 \le i \le n} \left| x_i \right| $ in eq. (\ref{e7})), 
for which their edge is equal to $r_{max}$, since distances greater then 
$r_{max}$ are not
considered in eq. (\ref{e7}). If, for example, one wants to calculate the
correlation of $r \le r_{max}=\frac{1}{10}D$ (in accordance with 
conditions 
(\ref{e10}) (\ref{e11}) (\ref{e13}) (\ref{e14})), then (in $2D$ projection)
it is enough to calculate distances in 9 squares instead 100 squares 
(actually, it is enough to consider just 5 squares since eq. (\ref{e7}) is
symmetric). According to Grassberger, one has to build a matrix for which any
cell in it points to the list of data points located in it. It is possible to
generalize this method to a $3D$ projection matrix.

The generalization to higher dimensions can be done very easily according to 
the string sort method. As a first step one has to prepare a string $y_i$ 
(\ref{e16}), which is the same operation as gridding the attractor by a 
hyper-grid of size $r$. The next step is to sort strings $s_i$ (\ref{e17}). 
For each reference point in eq. (\ref{e12}) the boxes adjoining the 
box with the reference point in it should be found. Now, it is left to find 
the distances between
the reference point to other points in neighboring boxes (we keep the pointer 
vector from string $y_i$ to the initial series and vice versa), and calculate 
the correlation  according to eq. (\ref{e12}). 

The algorithm described above is especially efficient for very complex systems 
with high FD. For example, for EEG series produce by brain activity, it well
known (except for very special cases such as epilepsy \cite{Babloyantz88})
that the FD is, at least, four. In this case $n_{min}=4$, and instead
of calculating distances in $l^4$ boxes ($l$ is the ratio between the attractor
diameter and $r$) it is sufficient to calculate distances in $3^4=81$ boxes. 
If, for example, $l=9$, just $3^4/9^4=\frac{1}{81}$ distances must be computed 
(or even less if one take into account also the symmetry of eq. (\ref{e12})).
In this way, despite the $O( N \log_2 N)$ operations needed for the initial 
sort, one can reduce significantly the computation time.

\subsection {Testing of GID algorithm and number of data points needed for 
calculating GID}

Let us test the algorithm of GID on random numbers. We create a random series
between 0 to 1 (uniform distribution). We then reconstruct the phase space
according to Takens theory (\ref{e1}). For any embedding dimension that we 
would like to reconstruct, the phase space we will get a dimension that
is the same as the embedding dimension, since the new reconstructed vectors
are random in the new phase space, and hence fill the space 
(in our case, a hyper-cube of edge 1).

For embedding dimension 1 (a simple line), if the edge length is $\varepsilon$,
then the probability to fall in any edge is also $\varepsilon$. The number of 
edges $\varepsilon$ that covers the series range $[0,1)$ is $\lfloor \frac
{1}{\varepsilon} \rfloor$. The probability to fall in the last edge 
is $ 1 \% \varepsilon$ (the residue of the division). For
generalization $q$ one gets :
\begin{equation} \label{e18}
\sum_{i=1}^{M(\varepsilon)} p_{i}^q = \lfloor 1/\varepsilon \rfloor
\varepsilon^q + (1 \% \varepsilon)^q.
\end{equation}
For embedding dimension 2, one has to square (\ref{e18}) since the probability
distribution on different sides is equal. 
For embedding dimension $n$ eq. (\ref{e2}) becomes :
\begin{equation} \label{e19}
I_q(\varepsilon) = \frac{1}{1-q} \ln [ \lfloor 1/\varepsilon \rfloor
\varepsilon^q + (1 \% \varepsilon)^q]^n.
\end{equation}
Opening eq. (\ref{e19}) according to the binomial formula will give all 
different
combinations for the partly contained boxes. Notice that this grid location 
fulfills requirement (\ref{e5}) for the minimum information function.
\begin{figure}[thb]
\psfig{figure=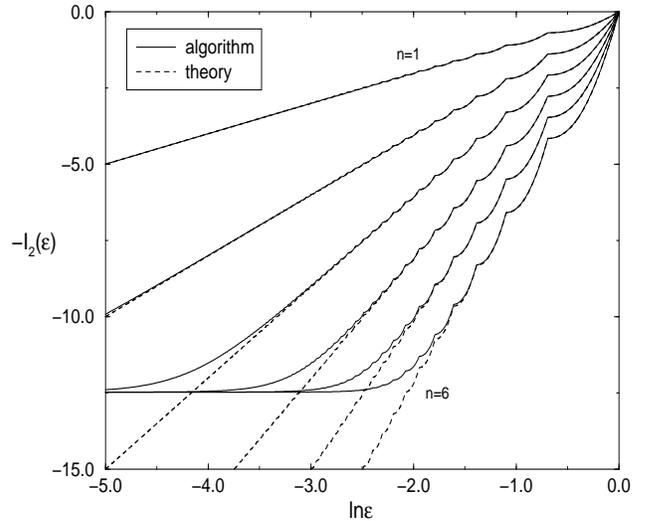,height=8cm,width=8.5cm,angle=-90}
\caption[]{\label{fig2}
Dimension calculation for random series by GID algorithm and by the theory.}
\end{figure}

In Fig. \ref{fig2} we present $-I_2(\varepsilon)$, both, 
according to the GID algorithm, and according to eq. (\ref{e19}). There is full
correspondence between the curves, and, as we find, 
it is possible to see that the slope
(dimension) in the embedding dimension $n$ is equal to the dimension itself. 
When $-I_2(\varepsilon) \approx -10$, the curves separate. There is lower 
convergence (when $-I_2(\varepsilon) \approx -12.5$) since the edge is too
small, and the number of reconstructed vectors is equal to the number of 
nonempty boxes.

The ``saw tooth'' that seen in Fig. \ref{fig2} caused by partial boxes that 
fall on the edge of the hyper-box. The ``saw tooth'' appears when 
there is an integer number of edges that covers the big edge. The slope in
the beginning of every saw tooth is equal to zero (that comes from the extremum
condition of (\ref{e19})). The ``saw teeth'' become smaller when 
$\varepsilon$ decrease, since the relative number of boxes that fall on the 
surface in small compared to the entire number of boxes.

It is possible to calculate roughly the number of points needed for the
dimension calculation in a homogeneous system. The separation point of
Fig. \ref{fig2} is 
approximately around $-I_2(\varepsilon) \approx -10$, for every embedding 
dimension. Thus, the number of hyper-boxes at this point is $e^{10}$ 
(since $-10=\ln \left(M(\varepsilon)p^2\right)$ where $p \sim 
\frac{1}{M(\varepsilon)}$ in our example). The 
graph come to a saturation around $-I_2(\varepsilon) \approx -12.5$, giving
rise to $e^{12.5}$ data points. The number of points per box is just, 
$e^{2.5} \approx 12$. Thus, generally, when there are less then 12 points per 
box, the value of $I_q(\varepsilon)$ is unreliable. Let us denote the minimum 
edges needed for the calculation as $m_{min}$ ($m_{min}$ is just 
$\lfloor 1/\varepsilon \rfloor$, and thus one can define a lower value for 
$\varepsilon$). The number of hyper-boxes in embedding dimension $n$ is 
$m_{min}^n$. Thus, the minimum number of points needed for computing the 
FD of an attractor with an FD $D$, is:
\begin{equation} \label{e20}
N = 12 m_{min}^D.
\end{equation}
This estimate is less then what was required by Smith \cite{Smit88} ($42^D$)
and larger then the requirement of Eckmann and Ruelle \cite{Eckm92} (if we
take, for example, $m_{min} = {\frac{1}{\rho}} = 10$ 
then according to (\ref{e10})
$10^{D/2}$ points is needed and according to (\ref{e20}) $12 \times 10^D$
points
is needed). However, as we pointed out earlier, (\ref{e20}) is a 
rough estimation, and one can converge to a desired FD even with less points.

\section {Examples} \label{s4}
In this section we will present the results of both GID and GCD on well known
systems, such as the Lorenz attractor and the Rossler attractor. 
The FD of those
results are well known by various methods, and we will present almost
identical results achieved by our methods. In the following examples, we 
choose one of the coordinates to be the time series; we normalize the time 
series to be in the range of $0 \ldots 1$.

\subsection {Lorenz Attractor} \label{s4a}
The Lorenz attractor \cite{Lorenz63} is represented by a set of three 
first order non-linear differential equations :
\begin{eqnarray}
\label{e21}
{dx \over dt} &= &\sigma(y-x) \nonumber \\
{dy \over dt} &= &-xz+rx-y \\ 
{dz \over dt} &= &xy-bz. \nonumber
\end{eqnarray}
Although it is much simpler to find the FD from the real phase-space $(x,y,z)$,
we prefer the difficult case of finding the FD just from one coordinate to 
show the efficiency of Takens theorem. We choose the $z^{th}$ component to be 
the time series from which we reconstruct the phase space (the parameter values
are: $\sigma=16$, $r=45.92$, $b=4$). We took $32768$ data points and $m=6$ (the
jumping on the time series which creates the reconstructed phase space; the 
time step is $\Delta t=0.02$). The generalization that we used was 
$q=2$. The embedding dimensions were $n=1..7$. The FD of Lorenz attractor 
is $2.07$ (\cite{Wolf85} and others).

\begin{figure}[thb]
\psfig{figure=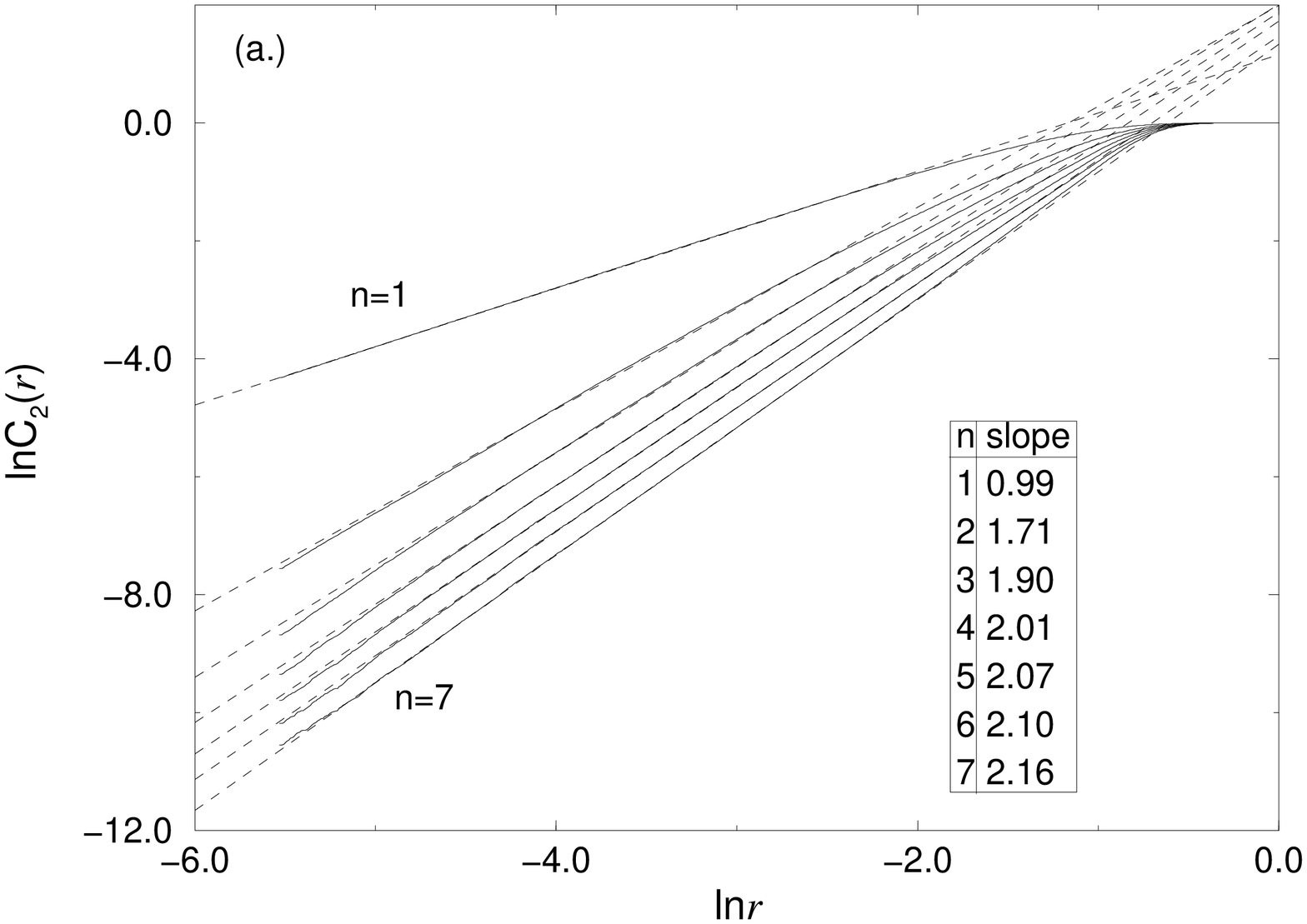,height=8cm,width=8.5cm,angle=-90}
\psfig{figure=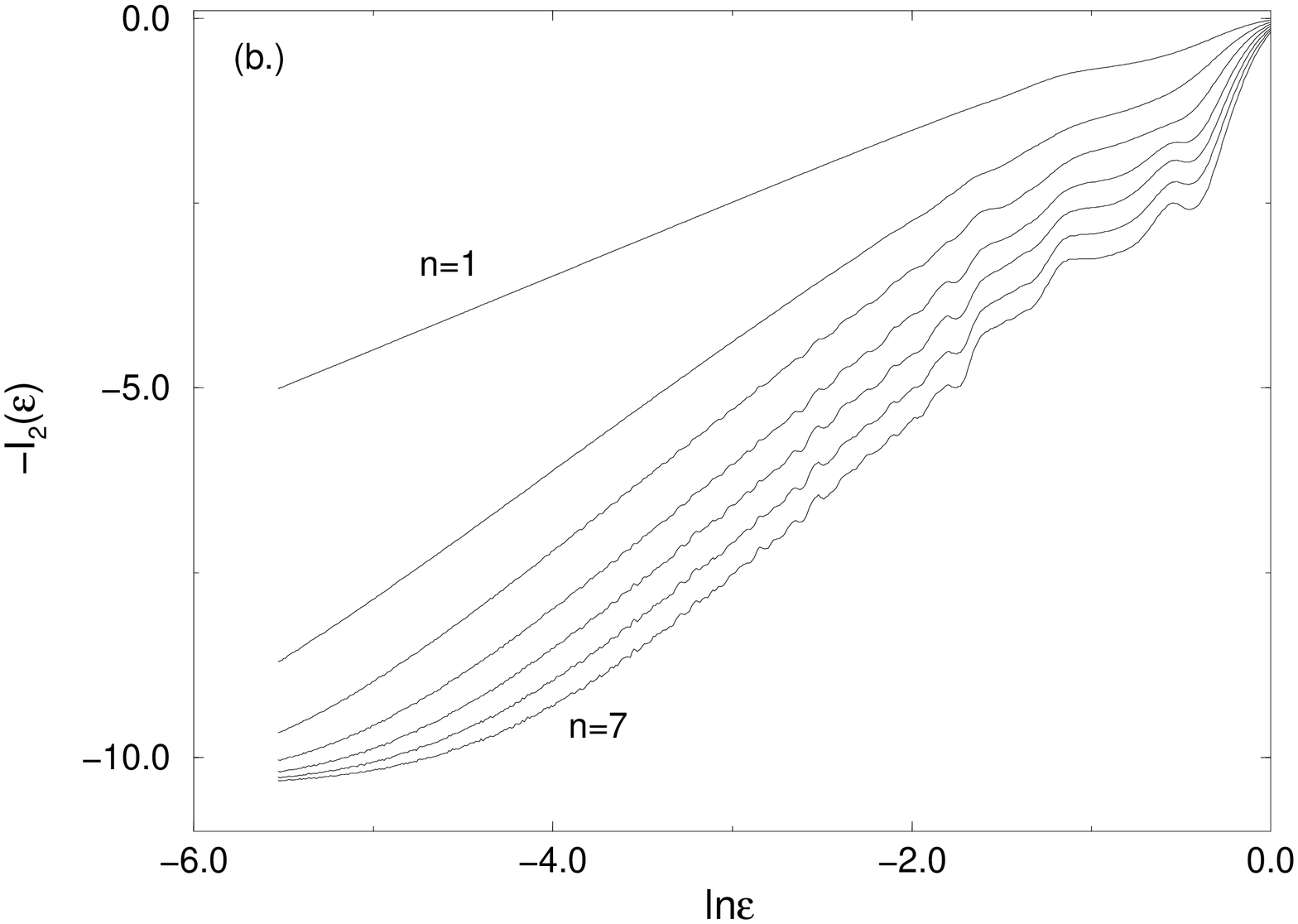,height=8cm,width=8.5cm,angle=-90}
\psfig{figure=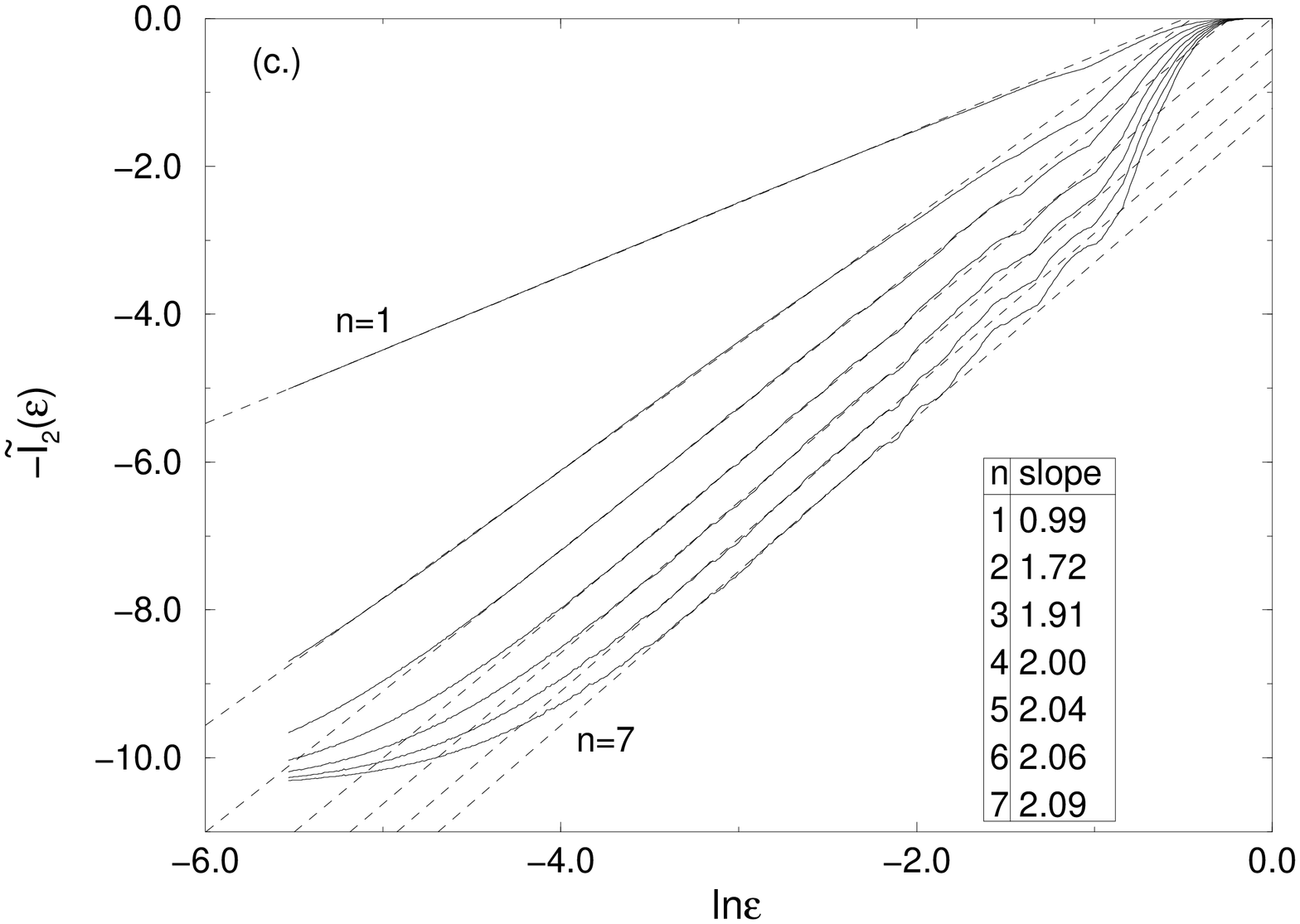,height=8cm,width=8.5cm,angle=-90}
\psfig{figure=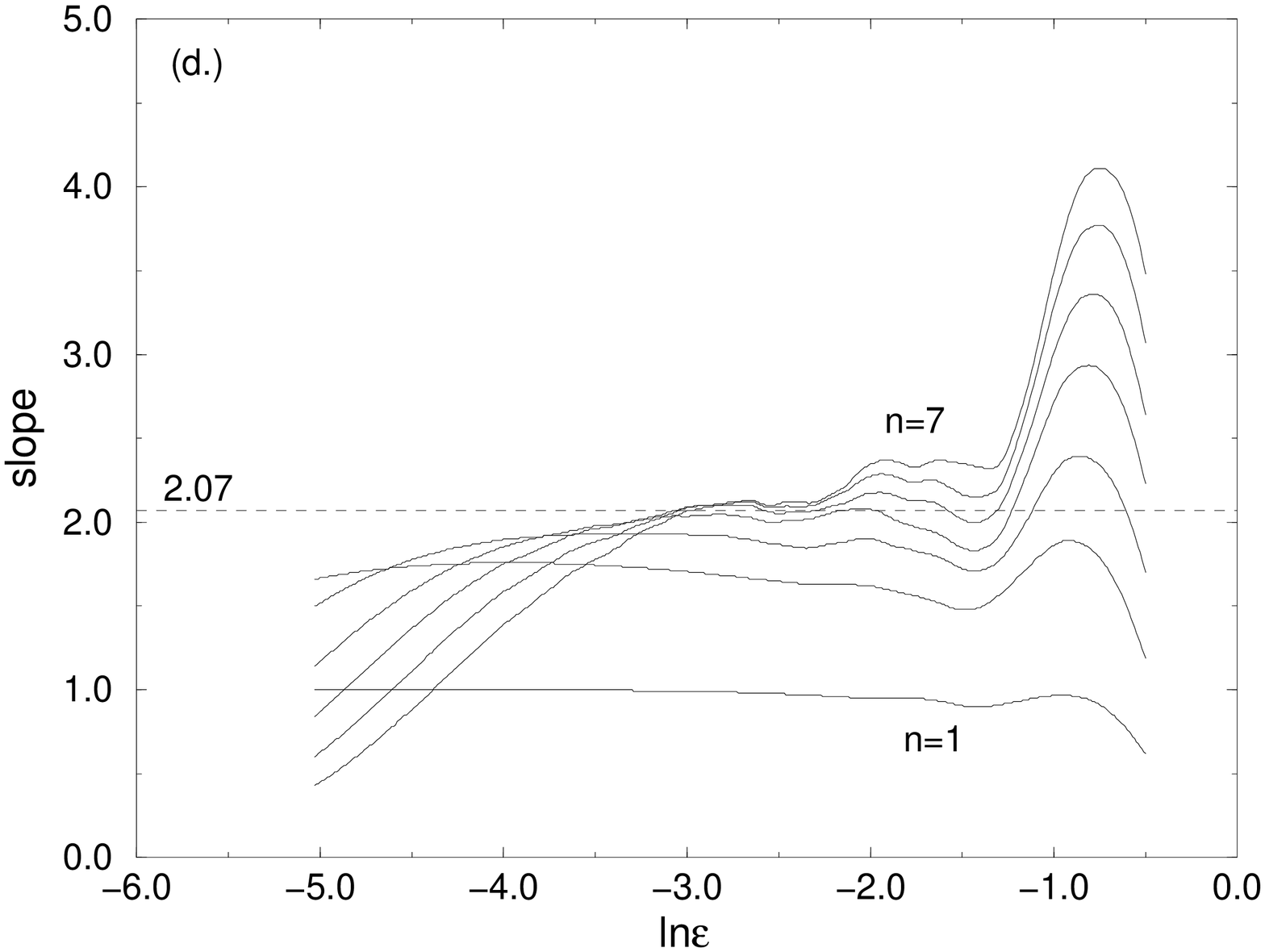,height=8cm,width=8.5cm,angle=-90}
\psfig{figure=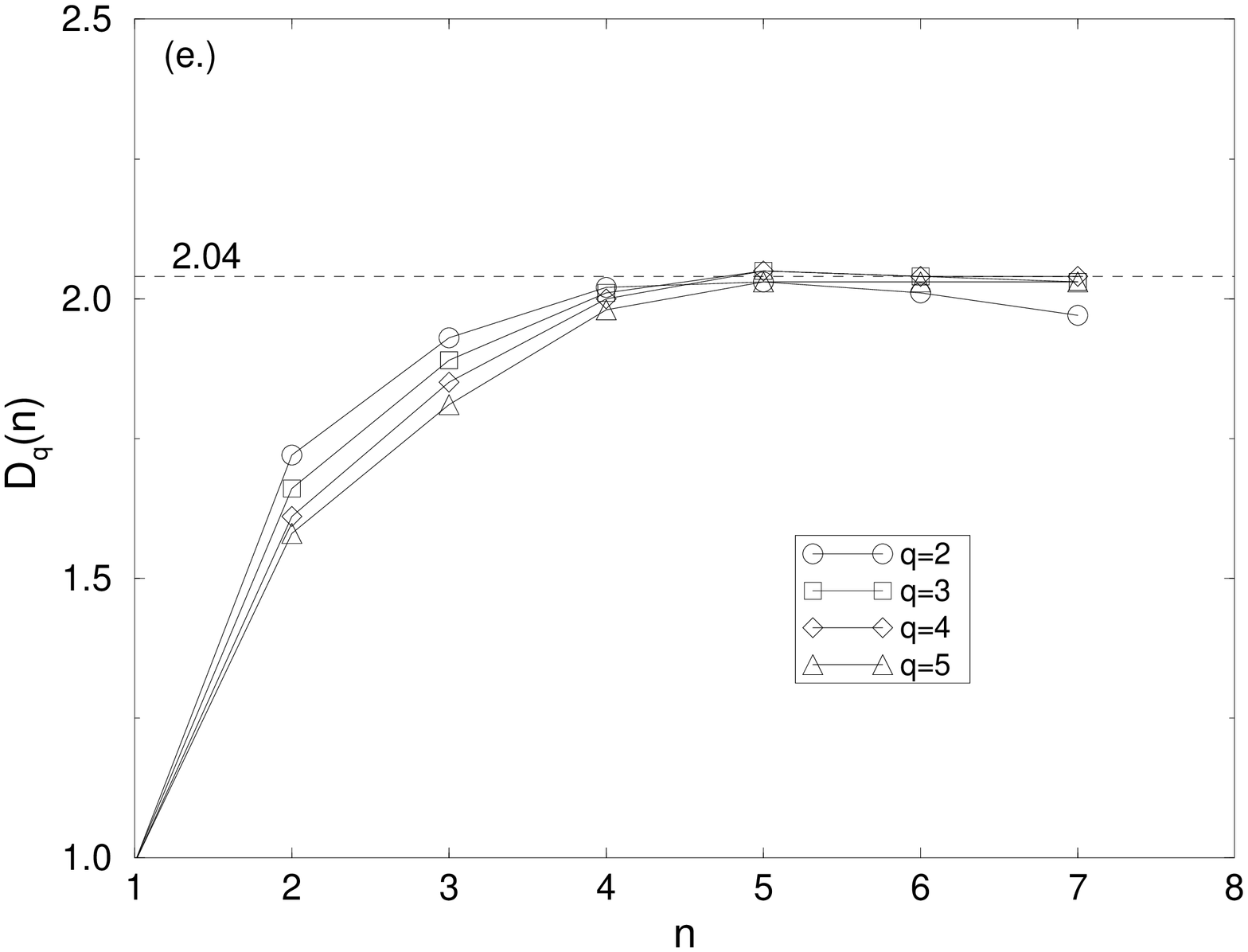,height=8cm,width=8.5cm,angle=-90}
\caption[]{\label{fig4}
Correlation and information curves of reconstructed Lorenz attractor.
a. Correlation curves; the estimate slopes are added.
b. Unsmoothed information curves.
c. Smoothed information curves; the estimated linear curves (dashed line) 
and their slope is added.
d. The slopes of $-\tilde I_2(\varepsilon)$ (Fig. \ref{fig4}c). 
The dashed line represents the estimated FD.
e. FD of Lorenz attractor for different embedding dimensions, $n$, and for 
different generalizations, $q$.
}
\end{figure}

In Fig. \ref{fig4} we show the results of GCD and GID methods. As
expected from the GCD curves, in Fig. \ref{fig4}a we see very smooth curves, 
for which the slopes
(in the center parts of the curves) converge approximately to dimension 
$2.08$. In Fig. \ref{fig4}b we see some jumping in the GID curves. It can be 
clearly seen that for small $\varepsilon$, one can not see the jumping while
for large $\varepsilon$ the non-monotonicity is much stronger. This is typical
behavior for GID curves; large $\varepsilon$ (or larger box edge) reflects more
variability of $-I_2(\varepsilon)$ to the hype- grid location. To smooth the
curves in order to get reliable results, it is necessary to find a proper 
location of the hyper-grid that gives minimum general information
$\tilde I_2(\varepsilon)$ (or maximum $-\tilde I_2(\varepsilon)$). 
We have performed 7 comparisons
to find that proper location. As for GCD, we also fitted approximate 
linear curves (dashed lines) which lead to dimension $2.09$ (Fig. \ref{fig4}c),
and thus, agree well with previous results. To make sure of the validity of the
dimension results we add here (Fig. \ref{fig4}d) graphs of successive slopes of
curves in Fig. \ref{fig4}c, and conclude with dimension results versus 
embedding dimension according to several generalizations ($q=2..5$, Fig. 
\ref{fig4}e). In both graphs, the approximate dimension is $2.03$. 

\subsection {Rossler Attractor} \label{s4b}
\begin{figure}[thb]
\psfig{figure=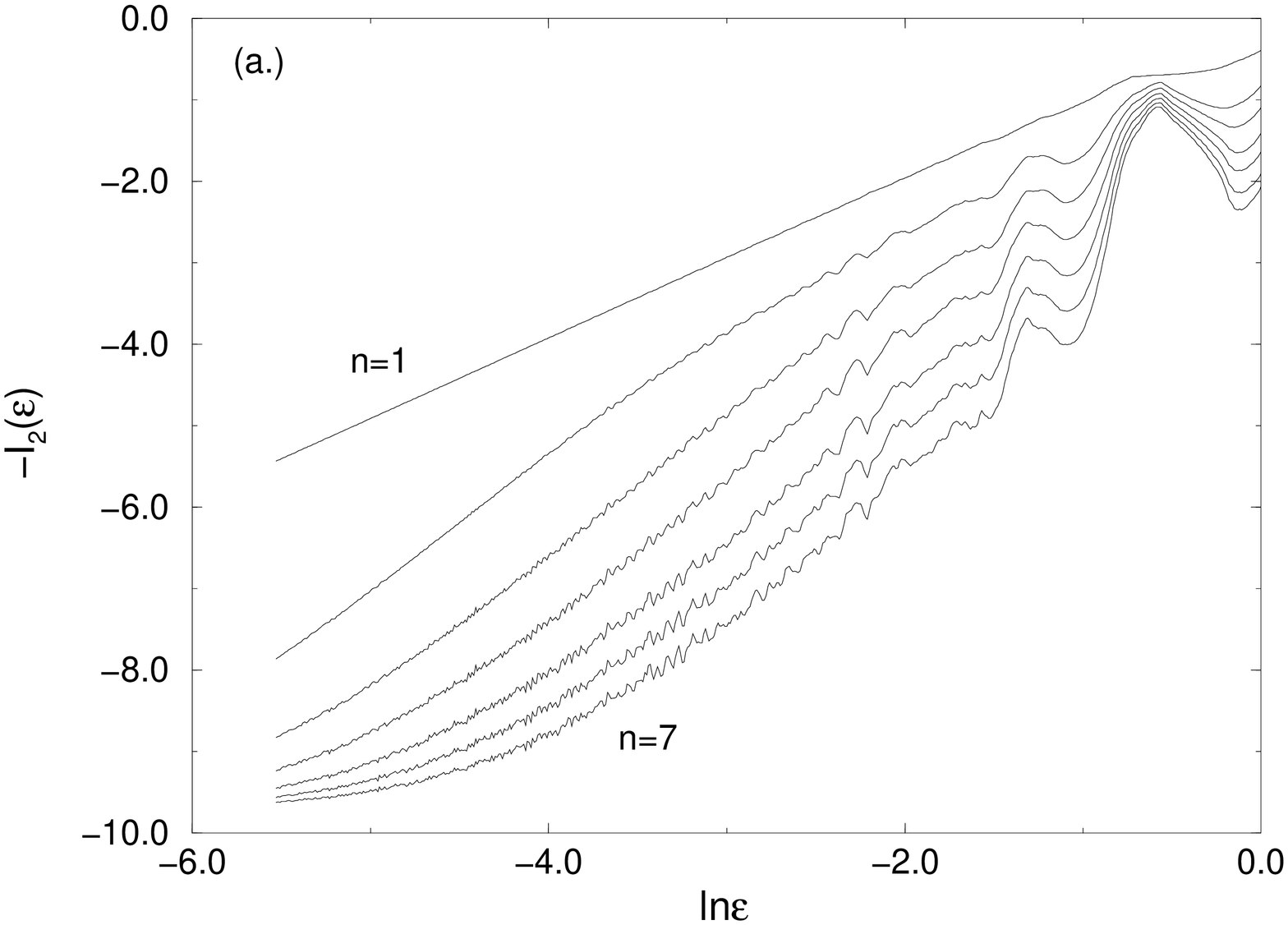,height=8cm,width=8.5cm,angle=-90}
\psfig{figure=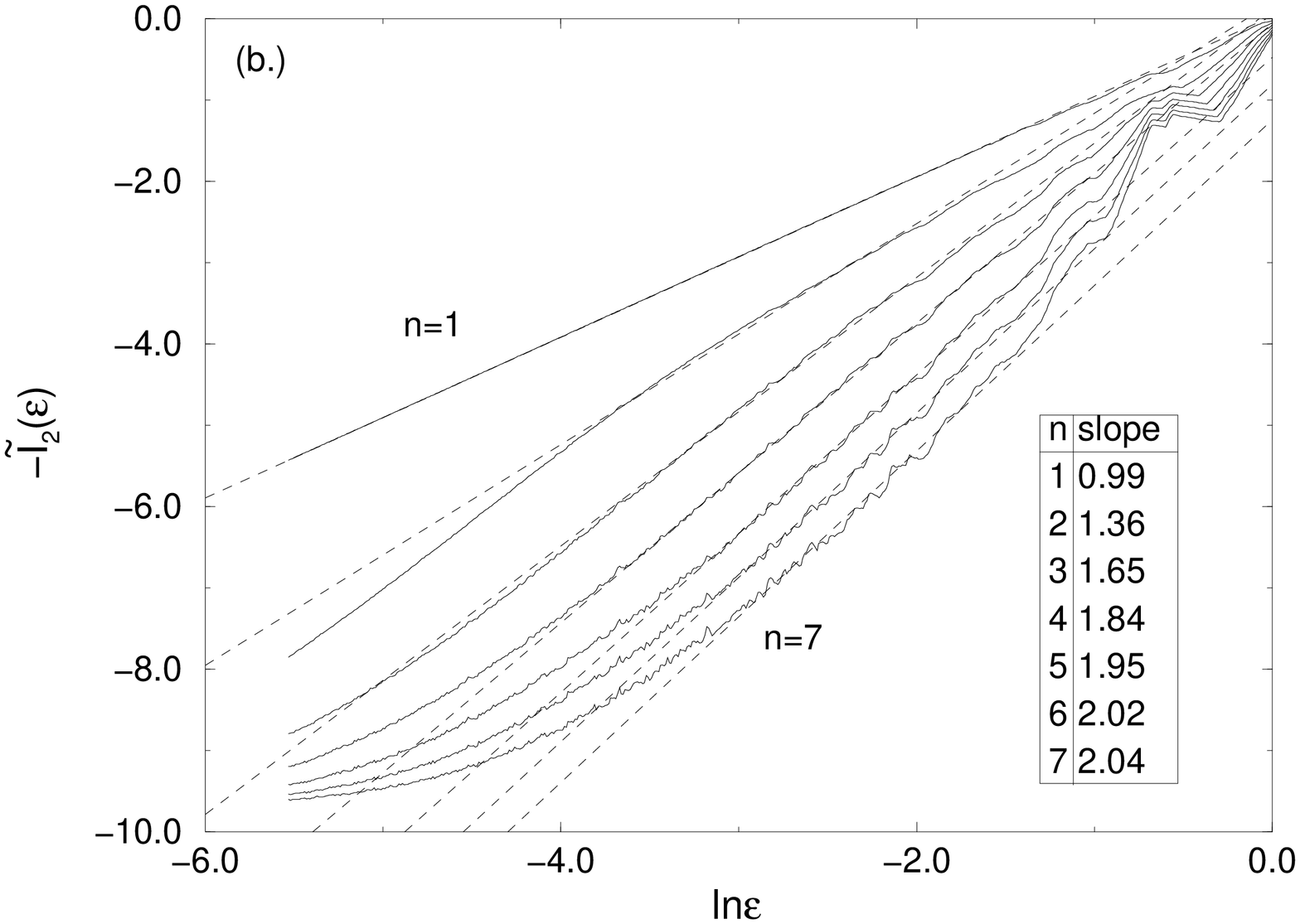,height=8cm,width=8.5cm,angle=-90}
\caption[]{\label{fig6}
a. Unsmoothed graphs of Rossler attractor.
b. Smoothed information graphs; the FD estimation is added.
}
\end{figure}

Let us examine briefly our second example, the Rossler attractor 
\cite{Rossler76}. The 
differential equations which generate the trajectory are :
\begin{eqnarray}
\label{e22}
\frac{dx}{dt} &= &-z-y \nonumber \\
\frac{dy}{dt} &= &x+ay\\
\frac{dz}{dt} &= &b+z(x-c). \nonumber
\end{eqnarray}
The parameter values are : $a=0.15$, $b=0.2$, $c=10$. The time step is $t=0.2$
and $m=8$. We use the $y$ direction to be the time series. The number of data
points is $N=16384$. In Fig. \ref{fig6} we show the results of GID method
of embedding dimension $n=1..7$. There are many more discontinuities in the 
unsmoothed curves (Fig. \ref{fig6}a) compared to the Lorenz attractor, 
reflecting a sensitivity to the grid location.
The smooth curves in Fig. \ref{fig6}b are produced after 20 comparisons. Again,
we approximate the slopes of central linear part of the curves in two different
ways and find dimension $2.05$, which is in agreement with dimension $2.03$ 
calculated by the use of the GCD method, and through the Lyapunov 
exponent \cite{Wolf85}.

\subsection {Van der Pol Oscillator} \label{s4c}
Another example that will be tested is the van der Pol oscillator
\cite{VanderPol}. This system
was investigated in the beginning of the century in order to find a model for
the heart beat variability (among other uses of this equation). The behavior
of the system (with parameter values that will be used) is chaotic although it 
looks quite periodic. Thus, we expect a smooth behavior of the geometrical 
structure of the attractor. The equation of motion is :
\begin{equation} \label{e23}
{\ddot x}-\alpha(1-x^2) \dot x +kx = f \cos \Omega t.
\end{equation}
The parameter values are : $\Omega=2.446$, $\alpha=5$, $k=1$ and $f=1$. Habib
and Ryne \cite{Habib95} and others \cite{Geist90} have found that the
Lyapunov exponents of the system are $\lambda_1 \approx 0.098$, $\lambda_2=0$ 
and $\lambda_3 \approx -6.84$,
and thus, according to the Kaplan and Yorke formula \cite{Kaplan79}, 
\begin{equation} \label{e24}
D_L=j+{\sum_{i=1}^j\lambda_i \over |\lambda_{j+1}|} \approx 2+
{0.098 \over 6.84} \approx 2.014
\end{equation}
where $j$ is defined by the condition that $\sum_{i=1}^{j} \lambda_i > 0$
and $\sum_{i=1}^{j+1} \lambda_i < 0$. The fact that the system has a 
``periodic'' nature is reflected in this very low
FD (as known, the minimal FD of a chaotic system is 2).
\begin{figure}[thb]
\psfig{figure=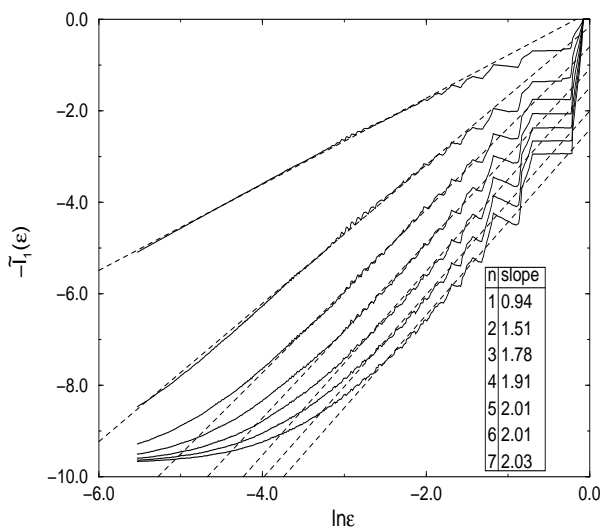,height=8cm,width=8.5cm,angle=-90}
\caption[]{\label{fig7}
van der Pol oscillator :
The smoothed information graphs, $-\tilde I_1(\varepsilon)$; The estimated 
FD is added (dashed lines with their corresponding slope).}
\end{figure}

In Fig. \ref{fig7} we present the calculation of $D_1$ which we suppose to be 
close to $D_L$. The approximate slope lead to an FD of $D_1 \sim 2.02$, which 
is in good agreement to $D_L$. Notice that minimization of $I_1(\varepsilon)$
did not succeed for large $\varepsilon$ values although we used 7 grid 
comparisons; that fact can cause a small error in the slope estimation, since
usually the slope is determined from the central part of the curves.

\subsection {Mackey-Glass Equation} \label{s4d}
Up to now we examined the relation between GID and GCD on systems with finite 
dimension, for which their FD can be calculated by using methods based on
the true integrated vector of the system. A class of systems which is more 
close to reality are systems with infinite dimensional dynamics. A delay
differential equation of the type
\begin{equation} \label{e25}
\frac{dx(t)}{dt}=F(x(t),x(t-\tau))
\end{equation}
belongs to this class. In order to solve this kind of equation, one needs to
initiate $x(t)$ in the range $0 \le t \le \tau$, and then, by assuming that the
system can be represented by a finite set of discrete values of $x(t)$, it is
possible to evaluate the system dynamics step by step. The solution is 
considered
to be accurate if one converges to the same global properties, such as, 
Lyapunov exponents and FD, by different methods \cite{Farmer81}, 
and without dependence on the number of intervals to which the function is
divided \cite{Gras83}\footnote{Note that the dynamics which is calculated 
by the use of different methods, or, equivalently, by a different number of 
discrete integration intervals, does not have to behave identically. In fact, 
one can expect similar dynamics if the system is not chaotic, but, if the 
system is chaotic, it is sensitive to initial condition and to the integration 
method, and even moreover, it is sensitive to integration accuracy 
\cite{{Greene78},{Ashkenazy98}}. However, the global properties of a system 
converge to the same values, by using different integration methods and using
different integration accuracy \cite{Ashkenazy98}.}. Delay equations, such as 
eq. (\ref{e25}), describe systems in which a stimulus has a delay response.
There are many practical examples from control theory, economics, population 
biology, and other fields.

One of the known examples is the model of blood cell production in patients 
with leukemia, formulated by Mackey and Glass \cite{Mackey77} :
\begin{equation} \label{e26}
\dot x(t)=\frac{ax(t-\tau)}{1+[x(t-\tau)]^{c}}-bx(t).
\end{equation}
Following refs. \cite{{Farmer81},{Gras83}}, the parameter values are : 
$a=0.1$, $b=0.2$, and $c=10$. We confine ourselves to $\tau=30.0$. 
As in ref. \cite{Gras83}, we choose the time series to be 
$\{x(t), x(t+\tau), x(t+2\tau), \ldots\}$\footnote{In fact, the common 
procedure of reconstructing a phase space from a time series is described in 
eq. (\ref{e1}). According to this method, one has to build the reconstructed 
vectors by the use of a jumping choice $m$ which is determined by the first 
zero
of the autocorrelation function, or the first minimum of the mutual information
function \cite{Sch89}. However, we used the same series as in 
ref. \cite{Gras83} in order to compare results.}, as well as the integration
method which is described in this reference. The length of the time series is
$N=131072$.
\begin{figure}[thb]
\psfig{figure=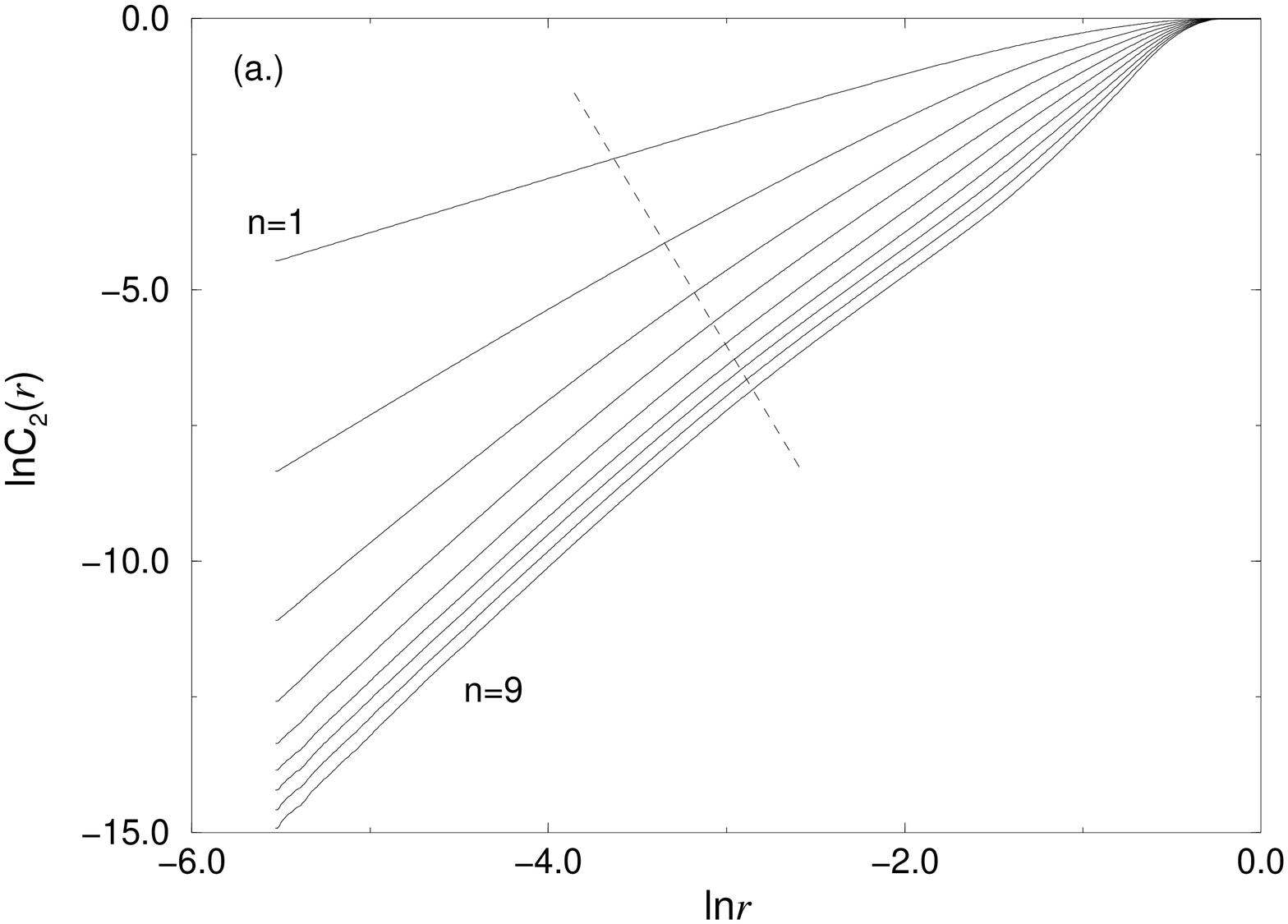,height=8cm,width=8.5cm,angle=-90}
\psfig{figure=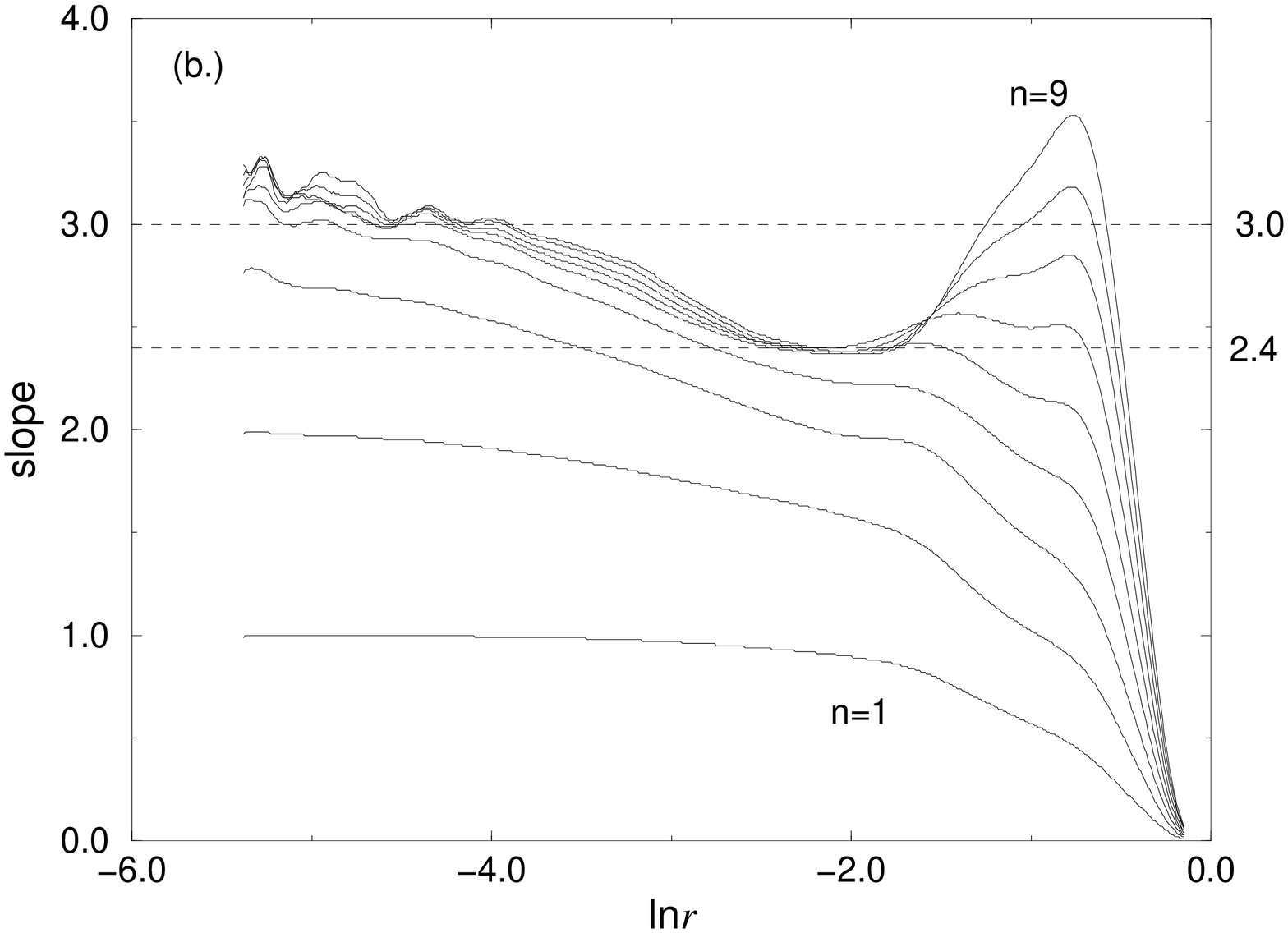,height=8cm,width=8.5cm,angle=-90}
\psfig{figure=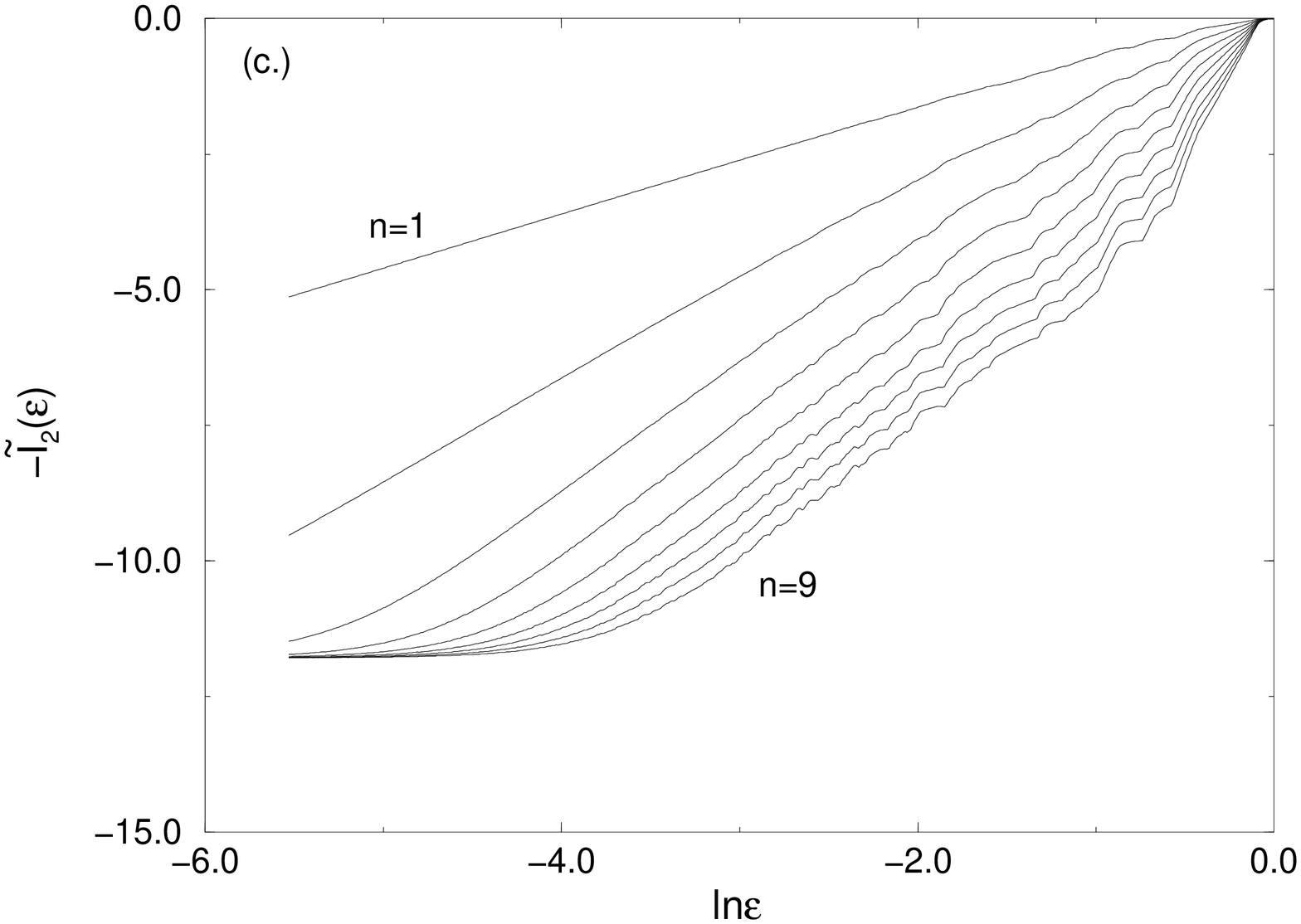,height=8cm,width=8.5cm,angle=-90}
\psfig{figure=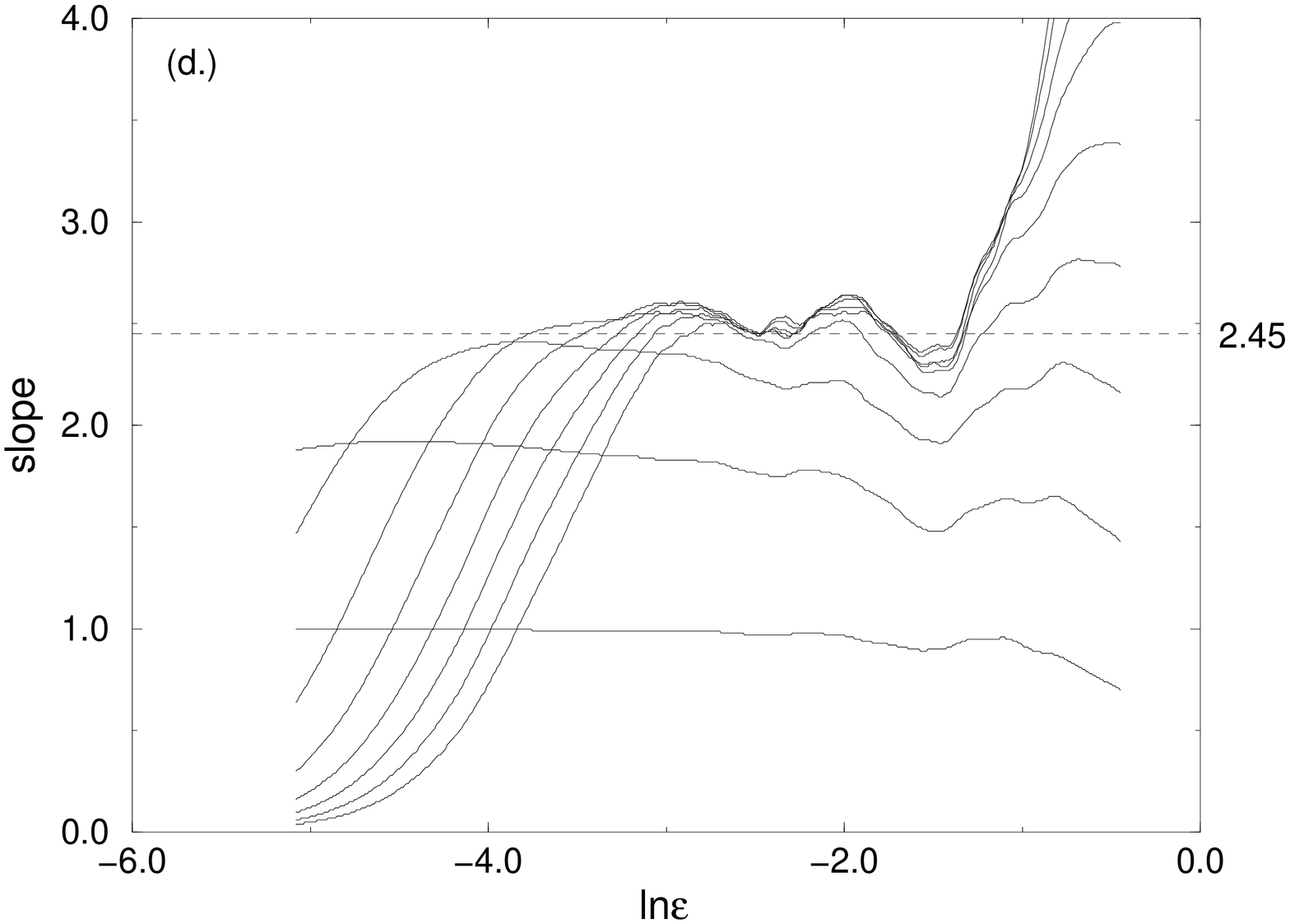,height=8cm,width=8.5cm,angle=-90}
\caption[]{\label{fig8}
Mackey-Glass equation: 
a. The GCD graphs. One notices two regions of parallel curves, which lead
to different FD's.
b. The successive slopes of a.
c. The GID graphs. 
d. The successive slopes of c.}
\end{figure}

The FD ($D_2$) calculation of the Mackey-Glass equation is presented in 
Fig. \ref{fig8}. The embedding dimensions are, $n=1 \ldots 9$. 
In Fig. \ref{fig8}a, the correlation function, $C_2(r)$ is 
shown. One notices that there are two regions in each one of which 
there is convergence to a certain slope. These regions are separate by a 
dashed line. In Fig. 
\ref{fig8}b, the average local slope of Fig. \ref{fig8}a is shown. One can 
identify easily two convergences, the first in the neighborhood of 
$\sim 3.0$, and the second around $\sim 2.4$. Thus, there are two 
approximations for the FD, $D_2$, pointing to two different scales. The first 
approximation is similar to the FD that was calculated in ref. \cite{Gras83}.
However, the GID graphs which are presented in Fig. \ref{fig8}c and d lead to 
an FD, $D_2 \sim 2.45$, which seems to be very close to the second convergence
of Fig. \ref{fig8}b. Notice that the convergence to $D_2 \sim 2.4$ appears,
in both methods, in the neighborhood of the same box size ($\sim 2.1$).

\section {Summary} \label{s6}

In this work we develop a new algorithm for calculation of a general 
information, $I_q(n)$, which is based on strings sorting (the method of string 
sort can be used 
also to calculate the conventional GCD method). According to our algorithm, 
one can divide the phase space into
255 parts in each hyper-box edge. The algorithm requires $O(N\log_2N)$ 
computations, where $N$ is the number of reconstructed vectors. A rough 
estimate for the number of points needed for the FD calculation was given. 
The algorithm, which can 
be used in a regular system with known equations of motion, was tested
on a reconstructed phase space (which was built according to Takens theorem).
The general information graphs have non monotonic curves, which can be smoothed
by the requirement for minimum general information. We examine our algorithm 
on some well known examples, such as, the Lorenz attractor, the Rossler 
attractor, the
van der Pol oscillator and others, and show that the FD that was computed by 
the GID method is almost identical to the well known FD's of those systems. 

In practice, the computation time of an FD using the GID method, was much less 
than for the GCD method. For a typical time series with $32768$ data points, 
the computation time needed for the GCD method was about nine times greater 
than the 
computation time of GID method (when we do not restrict ourselves to small
$r$ values and we compute all $N^2$ relations). Thus, in addition to the fact 
that the algorithm developed in 
this paper enables the use of comparative methods (which is crucial in some 
cases), the algorithm is generally faster.

The author wishes to thank J. Levitan, M. Lewkowicz and L.P. Horwitz for very 
useful discussions.


\vspace{-0.7truecm}

\begin{references}

\vspace{-1.8truecm}

\bibitem{Kaplan79} J.L. Kaplan and J.A. Yorke, in: {\it Functional
Differential Equations and Approximations of Fixed Points}, Lecture Notes
in Mathematics {\bf 730}, eds. H.-O. Peitgen and H.-O. Walther
(Springer Verlag, Berlin, 1979).

\bibitem{Babloyantz85} A. Babloyantz, C. Nicolis, and M. Salazar, Phys. Rev. A
{\bf 111}, 152 (1985); A. Babloyantz and A. Destexhe, Proc. Natl. Acad. Sci.
USA {\bf 83} 3513 (1986). 

\bibitem{Basar90} E. Basar, ed., {\it Chaos in the Brain} 
(Springer, Berlin, 1990).

\bibitem{MayerKress88} G. Mayer-Kress, F.E. Yates, L. Benton, M. Keidel, W.S. 
Tirsch and S.J. Poepple, Math. Biosci. {\bf 90}, 155 (1988).

\bibitem{Saermark89} K. Saermark, J. Lebech, C.K. Bak, and A. Sabers, in: 
Springer Series in Brain Dynamics, 2 eds. E Basar and T.H. Bullock (Springer,
Berlin, 1989) p. 149.

\bibitem{Saermark97} K. Saermark, Y. Ashkenazy, J. Levitan, and M. Lewkowicz,
Physica A {\bf 236}, 363 (1997).

\bibitem{Takens81} F. Takens, in : {\it Dynamical Systems and Turbulence}, eds.
D. Rand and L.S. Young, (Springer Verlag, Berlin, 1981).

\bibitem{Sch89} Discussions about the choose of $\tau$ see for example : 
H.G. Schuster, {\it Deterministic Chaos}, (Physik Verlag, Weinheim, 1989); 
A.M. Fraser and H.L. Swinney, Phys. Rev. A {\bf 33}, 1134 (1986); 
J.-C. Roux, R.H. Simoyi and H.L. Swinney, Physica D {\bf 8}, 257 (1983).

\bibitem{Abarbanel93} H.D.I. Abarbanel, R. Brown, J.J. Sidorowich, 
and L.S. Tsimring, Rev. Mod. Phys. {\bf 65}, 1331 (1993).

\bibitem{Ding93} M. Ding, C. Grebobi, E. Ott, T. Sauer and J.A. Yorke, Physica
D {\bf 8}, 257 (1993).

\bibitem{Shannon49} C.E. Shannon and W. Weaver, {\it The Mathematical Theory
of Information}, (University of Ill. Press, Urbana, 1949).

\bibitem{Bala76} J. Balatoni and A. Renyi, in : {\it Selected Papers of A. 
Renyi}, ed. P. Turan, (Akademiai K. Budapest, 1976), Vol. 1, p. 588.

\bibitem{Gras83} P. Grassberger and I. Procaccia, Physica D {\bf 9},
189 (1983).

\bibitem{Pawe87} K. Pawelzik and H.G. Schuster, Phys. Rev. A {\bf 35},
481 (1987).

\bibitem{Gras90} P. Grassberger, Phys. Lett. A {\bf 148}, 63 (1990).

\bibitem{Babloyantz88} A. Babloyantz and A. Destexhe, in: {\it From Chemical
to Biological Organization}, eds. M. Markus, S. Muller, and G. Nicolis 
(Springer Verlag, Berlin, 1988).

\bibitem{Smit88} L.A. Smith, Phys. Lett. A {\bf 133}, 283 (1988).

\bibitem{Eckm92} J.-P. Eckmann and D. Ruelle, Physica D {\bf 56},
185 (1992).

\bibitem{Thei86} J. Theiler, Phys. Rev. A {\bf 34}, 2427 (1986).

\bibitem{Casw86} W.E. Caswell and J.A. York, in: {\it Dimensions and
Entropies in Chaotic Systems}, ed. G. Mayer-Kress, 
(Springer Verlag, Berlin, 1986).

\bibitem{Molteno93} T.C.A. Molteno, Phys. Rev. E {\bf 48}, 3263 (1993).

\bibitem{Hou90} X.-J. Hou, R. Gilmore, G. Mindlin, and H. Solari, Phys. Lett. 
A {\bf 151}, 43 (1990).

\bibitem{Block90} A. Block, W. von Bloh, and H. Schnellnhuber, Phys. Rev. A 
{\bf 42}, 1869 (1990).

\bibitem{Lieb89} L. Liebovitch and T. Toth, Phys. Lett. A {\bf 141}, 386 
(1989).

\bibitem{Kruger96} A. Kruger, Comp. Phys. Comm. {\bf 98}, 224 (1996).

\bibitem{nr95} W.H. Press, S.A. Teukolsky, W.T. Vetterling, and B.P. Flannery,
{\it Numerical Recipes in C}, 2nd edn. (Cambridge University, Cambridge, 1995).

\bibitem{Sort} D.E. Knooth, {\it The Art of Computer Programming} vol. 3 
- Sorting and Searching, (Addison-Wesley, Reading, Mass., 1975); 
There is other sort algorithms that require even less computation 
($O(N)$) such as radix sorting, A.V. Aho, J.E. Hopcroft, and J.D. Ullman,
{\it Data Structures and Algorithms}, (Addison-Wesley, Mass., 1983).

\bibitem{Lorenz63} E.N. Lorenz, J. Atmos. Sci. {\bf 20}, 130 (1963).

\bibitem{Wolf85} A. Wolf, J.B. Swift, H.L. Swinney and J.A. Vastano,
Physica D {\bf 16}, 285 (1985).

\bibitem{Rossler76} O.E. Rossler, Phys. Lett. {\bf 57A}, 397 (1976).

\bibitem{VanderPol} B. van der Pol, Phil. Mag. (7) {\bf 2}, 978 (1926);
B. van der Pol and van der Mark, Phil. Mag. (7) {\bf 6}, 763 (1928).

\bibitem{Habib95} S. Habib and R.D. Ryne, Phys. Rev. Lett. {\bf 74},
70 (1995).

\bibitem{Geist90} K. Geist, U. Parlitz, and W.Lauterborn, Prog. Theor. Phys.
{\bf 83}, 875 (1990).

\bibitem{Farmer81} J.D. Farmer, Physica D {\bf 4}, 366 (1981).

\bibitem{Greene78} J.M. Greene, J. Math. Phys. {\bf 20}, 1183 (1979).

\bibitem{Ashkenazy98} Y. Ashkenazy, C. Goren, and L.P. Horwitz, Phys. Lett.
A, {\bf 243}, 195 (1998).

\bibitem{Mackey77} M.C. Mackey and L. Glass, Science {\bf 197}, 287 (1977).

\end{references}
\end{document}